# ON THE PREDICTION OF VELOCITY FIELDS FROM REDSHIFT SPACE GALAXY SAMPLES

**Adi Nusser** and **Marc Davis**

Center for Particle Astrophysics and Astronomy Department

University of California, Berkeley, CA 94720

## ABSTRACT

We present a new method for recovering the underlying velocity field from an observed distribution of galaxies in redshift space. The method is based on a kinematic Zel'dovich relation between the velocity and density fields in redshift space. This relation is expressed in a differential equation slightly modified from the usual Poisson equation, and which depends non-trivially on $\beta \equiv \Omega^{0.6}/b$. The linear equation can be readily solved by standard techniques of separation of variables by means of spherical harmonics. One can also include a term describing the "rocket effect" discussed by Kaiser (1987). From this redshift space information alone, one can generate a prediction of the peculiar velocity field for each harmonic $(l, m)$ as a function of distance. We note that for the quadrupole and higher order moments, the equation is a boundary value problem with solutions dependent on both the interior and exterior mass distribution. However, for a shell at distance $r$, the dipole, as well as the monopole, of the velocity field in the Local Group frame is fully determined by the interior mass distribution. This implies that the shear of the measured velocity field, when fit to a dipole distortion, should be aligned and consistent with the gravity field inferred from the well determined local galaxy distribution. As a preliminary application we compute the velocity dipole of distant shells as predicted from the 1.2Jy IRAS survey compared to the measured velocity dipole on shells, as inferred from a recent POTENT analysis. The coherence between the two fields is good, yielding a best estimate of $\beta = 0.6 \pm 0.2$.

# 1. INTRODUCTION

The observed distribution of galaxies in redshift space may differ significantly from the distribution of the underlying matter in real space (e.g, Kaiser 1987). This is due to three effects: *a*) pure redshift distortions arising from the mapping from real to redshift space, *b*) redshift space distortions coupled to a steep selection function $\phi(r)$ which introduces false modes, the "rocket effect", and *c*) a possible biasing in the distribution of galaxies relative to the underlying matter.

In view of the accumulating large galaxy redshift and peculiar velocity surveys, it would be extremely worthwhile if one could recover the associated velocity and density fluctuation fields in real space. Yahil *et al.* (1991) describe a method in which the distribution of galaxies in real space is estimated by iterations under the constraint that the predicted real space velocity and density fields satisfy the relation given by linear theory. Here we develop analytical relations between the velocity and density in redshift space. This will allow a recovery of the velocity field directly in redshift space. Under the assumption of Zel'dovich displacements of particles we find linear and quasilinear approximations relating the velocity to the density in redshift space. Unfortunately, a quasilinear approximation in redshift space is difficult to use for recovering the velocity from density. The complication arises because the flow in redshift space is not vorticity free in the quasilinear regime even for a potential flow in real space. However, in the linear regime the property of potential flow is conserved in both spaces. Thus we gain a great simplification by retaining the analysis strictly for linear perturbations. Here we shall use the quasilinear approximation only for deriving the linear relation by maintaining only first order terms. However, one should keep in mind that in regions like the Great Attractor and the large voids, the overdensity is of order unity, and nonlinear corrections can be important.

In §2 we present the analytical formalism and develop linear and quasilinear approximations in redshift space, and discuss implementing the linear velocity-density relation to recover the velocity from flux-limited redshift surveys. In §3 we bring a preliminary application of the method to recover the velocity dipole from the 1.2Jy IRAS galaxies. This gravitational dipole is then compared to the measured dipole, from which one can estimate the cosmological parameter $\beta = f(\Omega)/b$. We conclude by the discussion in §4.

## 2. THE FORMALISM

### 2.1 Pure Redshift Distortions

Assume we are given the underlying density field in redshift space, deferring for the next subsection how it could be derived from observations. We wish to establish the relationship between the velocity and density fields; all in redshift space. We shall measure all redshifts in the LG frame.

Let $\boldsymbol{v}$, $\boldsymbol{q}$, $\boldsymbol{x}$ and $\boldsymbol{s}$ be, respectively, the comoving peculiar velocity, the initial coordinate (*i.e*, Lagrangian), the present real space coordinate (*i.e*, Eulerian), and the redshift space coordinate of a particle (galaxy) in units of km/s; the corresponding physical quantities are $a\boldsymbol{v}$, $a\boldsymbol{q}$, $a\boldsymbol{x}$ and $a\boldsymbol{s}$, where $a(t)$ is the scale factor. The comoving redshift and real coordinates are related by

$$\boldsymbol{s} = (x + (U - U_0))\hat{\boldsymbol{x}} = \boldsymbol{x} + (U - U_0)\hat{\boldsymbol{x}} \qquad (1)$$

where $\hat{\boldsymbol{x}} = \boldsymbol{x}/x$ and $U = \boldsymbol{v} \cdot \hat{\boldsymbol{x}}$ is the radial component of the peculiar velocity and $U_0 = \boldsymbol{v}(0) \cdot \hat{\boldsymbol{x}}$ assuming the observer comoving with the flow of matter at the Local Group (LG) which we chose to be at $\boldsymbol{x} = 0$. This relation is independent of dynamics. To describe dynamics we adopt the Zel'dovich approximation (Zel'dovich 1970). Provided the motion of galaxies is not severely affected by nonlinear effects, the Zel'dovich approximation is an excellent tool for describing their dynamical evolution (Nusser *et al.* 1991). Unlike the pure linear approximation, the Zel'dovich approximation incorporates the displacement of particles from their initial positions. In order for the analysis presented here to apply we shall make the assumption of no orbit-mixing either in real or in redshift space, *i.e.*, there is a one to one correspondence between $\boldsymbol{x}$ and $\boldsymbol{q}$ and between $\boldsymbol{s}$ and $\boldsymbol{q}$. This requirement is enforced in practice by smoothing all fields over a sufficiently large scale. The smoothing procedure is also essential to avoid severe nonlinear effects.

According to the Zel'dovich approximation, the Eulerian coordinate of a particle relative to the LG ($\boldsymbol{x} = 0$) is,

$$\boldsymbol{x} = \boldsymbol{q} + \frac{1}{f(\Omega)}\boldsymbol{v}, \qquad (2)$$

where $f(\Omega) \approx \Omega^{0.6}$. Therefore, unless $\Omega \ll 1$, the mapping (1) from real to redshift coordinates is comparable to the dynamical displacement from Lagrangian to Eulerian coordinates. In fact for $\Omega = 1$ the redshift radial coordinate of a particle equals its real radial coordinate at twice the expansion time.

Writing all quantities in terms of the redshift space coordinate, (1) and (2) yield

$$\boldsymbol{q}(\boldsymbol{s}) = \boldsymbol{s} - \boldsymbol{F}(\boldsymbol{s}) - \frac{1}{f}\boldsymbol{v}(0), \qquad (3)$$

where

$$\boldsymbol{F}(\boldsymbol{s}) = \frac{1}{f}\boldsymbol{v}_{lg}(\boldsymbol{s}) + u\hat{\boldsymbol{s}},$$

$\boldsymbol{v}_{lg} = \boldsymbol{v} - \boldsymbol{v}(0)$, the comoving peculiar velocity in the LG frame, and $u = U - U(0)$, the radial component of the comoving peculiar velocity in the LG frame.

In writing the last equation we have used $\hat{\boldsymbol{s}} = \boldsymbol{s}/s = \hat{\boldsymbol{x}}$. In the absence of orbit-mixing, mass conservation reads

$$\rho_q \mathrm{d}^3 q = \rho(\boldsymbol{s})\mathrm{d}^3 s, \qquad (4)$$

where $\rho$ is the redshift space density and $\rho_q$ is the Lagrangian density, *i.e.*, the mean density of the background. By (3) the density contrast, $\delta = (\rho - \rho_q)/\rho_q$, is therefore approximated by

$$\delta(\boldsymbol{s}) = \left\| I - \frac{\partial \boldsymbol{F}}{\partial \boldsymbol{s}} \right\| - 1, \qquad (5)$$

where the double vertical bars denote the Jacobian determinant and $I$ is the unit matrix. This is a non-linear expression involving first order partial derivatives of $\boldsymbol{F}$ and hence first derivatives of $\boldsymbol{v}_{lg}$. This approximation is the counterpart to the quasi-linear real space velocity-density relation found by Nusser *et al.* 1991.

Expanding the Jacobian in (5) to first order in $\boldsymbol{F}$ yields the following linear approximation,

$$\delta(\boldsymbol{s}) = -\boldsymbol{\nabla} \cdot \boldsymbol{F}(\boldsymbol{s}) = -\frac{1}{f}\boldsymbol{\nabla} \cdot \boldsymbol{v}_{lg}(\boldsymbol{s}) - \boldsymbol{\nabla} \cdot [u(\boldsymbol{s})\hat{\boldsymbol{s}}], \qquad (6)$$

where here and thereafter all derivatives are performed in redshift space. This is different from the usual linear real space velocity-density relation by the presence of the radial velocity divergence. In general this additional term is comparable to the density contrast in real space which, to first order, is identified with the full velocity divergence term in (6). In the following we shall restrict the treatment to the linear approximation.

If we assume a potential flow in real space then, to first order, the velocity in redshift space is derived from a potential. Although the velocity in real space remains irrotational even when the flow is non-linear as long as orbit-mixing is absent, vorticity is introduced to the velocity field in redshift space from second order effects, which we shall here ignore.

Defining a velocity potential, $\Phi$, by $\boldsymbol{v}_{lg}(\boldsymbol{s}) = -\nabla\Phi(\boldsymbol{s})$ and referring to (6) we obtain

$$\frac{1}{f}\nabla^2\Phi + \frac{1}{s^2}\frac{\partial}{\partial s}\left(s^2\frac{\partial\Phi}{\partial s}\right) = \delta, \qquad (7)$$

This is the equivalent of Poisson equation in real space. The form of (7) suggests separating the angular and radial dependence of the potential. Hence we expand the potential and density in Spherical Harmonics. Thus if in (7) we write $\Phi(\boldsymbol{s}) = \sum_{l=0}^{\infty}\sum_{m=-l}^{l}\Phi_{lm}(s)Y_{lm}(\theta,\varphi)$ and similarly for $\delta$ we find that

$$\frac{1}{s^2}\frac{\mathrm{d}}{\mathrm{d}s}\left(s^2\frac{\mathrm{d}\Phi_{lm}}{\mathrm{d}s}\right) - \frac{1}{1+f}\frac{l(l+1)\Phi_{lm}}{s^2} = \frac{f}{1+f}\delta_{lm}. \qquad (8)$$

The decaying and growing solutions of this equation without the source term are, $s^{\tilde{l}}$, and, $s^{-(\tilde{l}+1)}$, with $l \geq \tilde{l} \geq 0$ satisfying $(1+f)\tilde{l}(\tilde{l}+1) - l(l+1) = 0$. Therefore, the general solution to (8) in terms of the radial velocity multipoles, $u_{lm} = -\mathrm{d}\Phi_{lm}/\mathrm{d}s$, is

$$u_{lm}(s) = -\frac{\frac{f}{1+f}}{2\tilde{l}+1}\left[(\tilde{l}+1)s^{-(\tilde{l}+2)}\int_{s_1}^{s}\delta_{lm}(a)a^{\tilde{l}+2}\mathrm{d}a + \tilde{l}s^{\tilde{l}-1}\int_{s_2}^{s}\delta_{lm}(a)a^{-(\tilde{l}-1)}\mathrm{d}a\right], \qquad (9)$$

where the constants, $s_1$, and, $s_2$, are fixed by the boundary conditions. Since we consider here the velocity field in the LG frame, the velocity must vanish at small distances. This requirement is sufficient to determine the dipole and the monopole of the velocity field, because for these multipoles $\tilde{l} \leq 1$ and the velocity vanishes at the origin if and only if $s_1 = s_2 = 0$. Therefore the monopole and dipole of the velocity in the LG frame at a shell a distance $s$ from us is determined uniquely by the distribution of matter inside that shell. It is easy to see that this is also true for the velocity dipole determined from the Poisson equation given the density fluctuations in real space.

Multipoles higher than the dipole vanish by definition at the origin, and since the solution of each mode must be everywhere finite, the appropriate choice of integration bounds is then $s_1 = 0$ and $s_2 = \infty$. In order to compute these multipoles it is therefore necessary to know the distribution of matter over all space, but in practice we note that the mass density of shells exterior to the shell in question is weighted as $s^{-(\tilde{l}-1)}$, so that the influence of very distant shells is small, especially for high multipoles. In any event, we have reasonable knowledge of the mass distribution of IRAS galaxies to a redshift of 20,000 km/s, which provides adequate buffer for the velocity field interior to 10,000 km/s.

## 2.2 Selection Effects and Biasing

In the previous subsection we had in mind an ideal situation in which we have a perfect volume-limited sample of unbiased distribution of galaxies. In flux-limited surveys such as the IRAS sample, the number density of observed galaxies is a decreasing function of distance. If the observed redshift space number density is $n_0(\boldsymbol{s})$, then the true number density of galaxies is $n(\boldsymbol{s}) = n_0(\boldsymbol{s})/\phi(|\boldsymbol{x}|)$, where $\phi$ is the selection function and $\boldsymbol{x}$ is the real space position of galaxies at a redshift $\boldsymbol{s}$. Again we assume no orbit-mixing. Hence, even if we are interested only in the distribution of galaxies in redshift space we need to know the mapping from redshift to real space. Approximating $\boldsymbol{x}$ by $\boldsymbol{s}$ in the selection function leads to systematic biases as was first discussed by Kaiser (1987). But if we write $\phi(x) = \phi(s - u) \approx \phi(s) + (\mathrm{d}\phi/\mathrm{d}s)u(\boldsymbol{s})$ we find that

$$n(\boldsymbol{s}) = \frac{n_0(\boldsymbol{s})}{\phi(s)}\left[1 + \frac{1}{s}\frac{\mathrm{d}\ln\phi}{\mathrm{d}\ln s}u(\boldsymbol{s})\right], \qquad (10)$$

and

$$\delta^g(\boldsymbol{s}) = \frac{n(\boldsymbol{s}) - \bar{n}}{\bar{n}} = \delta_0^g(\boldsymbol{s}) + \frac{1}{s}\frac{\mathrm{d}\ln\phi}{\mathrm{d}\ln s}u(\boldsymbol{s}) \qquad (11)$$

where $\delta_0^g$ is the fractional excess of $n_0(\boldsymbol{s})/\phi(s)$ which is unambiguously determined. The logarithmic derivative which generates the difference between $n_0(\boldsymbol{s})/\phi(s)$ and the desired $n(s)$ can be quite large. For instance it is $\sim -2$ at a distance $\sim 3000$ km s$^{-1}$ and $\sim -5$ at a distance of $\sim 10000$ km s$^{-1}$ for the 1.2Jy IRAS sample. Considering that large scale peculiar velocities can be as high as 1000 km s$^{-1}$ this effect might be very serious (e.g, Kaiser 1987, Strauss *et al.* 1992).

The large scale galaxy distribution may not be an honest tracer of the underlying mass fluctuations, i.e. galaxy formation may be biased. The biasing, if present, must be a function of time [1], but in the usual sense for linear theory analysis, its effects enter via the factor $\beta = f/b$. Including the bias and equation (11) into the linear relation (8), we have

$$\frac{1}{s^2}\frac{\mathrm{d}}{\mathrm{d}s}\left(s^2\frac{\mathrm{d}\Phi_{lm}}{\mathrm{d}s}\right) - \frac{1}{1+\beta}\frac{l(l+1)\Phi_{lm}}{s^2} = \frac{\beta}{1+\beta}\left(\delta^g_{0\,lm} - \frac{1}{s}\frac{\mathrm{d}\ln\phi}{\mathrm{d}\ln s}\frac{\mathrm{d}\Phi_{lm}}{\mathrm{d}s}\right). \qquad (12)$$

Unless the selection function is very steep at small distances, the velocity monopole and dipole are found by requiring the velocity to vanish at the origin. Indeed, if $\phi \propto s^{-p}$ and $p \leq (2+4\beta)/\beta$ then, the solutions to (13) with $l = 1$ and $l = 0$ are uniquely determined by specifying vanishing velocity at the origin. This criterion is well satisfied for all existing redshift surveys.

### 3. PRELIMINARY APPLICATION

As a first qualitative application of the method in order to demonstrate its effectiveness, we consider the velocity dipole predicted from the distribution of the 1.2Jy IRAS galaxies (Fisher 1992). We generate a smooth galaxian density field in redshift space by assigning a Gaussian window to each galaxy with width proportional to the inverse of the selection function at the galaxy redshift position. We integrate Equation (12) with $l = 1$ using a fourth order Runge-Kutta integrator for several values of $\beta$.

Figure 1 shows the scalar amplitude of the velocity dipole as a function of limiting shell depth. This dipole is a measure of the expected motion of a thin shell relative to the local group, with the first term in Equation (9) (for $l = 1$) expressing the motion of the shell generated by the inhomogeneous mass distribution interior to it, and the second term expressing the motion of the central observer generated by the totality of mass interior to the shell. Only the second term is present in the analysis of Strauss *et al.* (1992), so Figure 1 should be similar, but not identical to Figure 2b of Strauss *et al.* which used the iteration method of Yahil *et al.* (1991). The two calculations agree very well in direction and amplitude, as expected, since the motion of the most distant shells relative to the CMB frame is expected to be small.

---

[1] We point out that continuity leads to the conclusion that $b$ must be a function of time approaching unity as time goes by. For example, if $\Omega = 1$ then, $b(t) = [a_0/a(t)](b_0 - 1) + 1$.

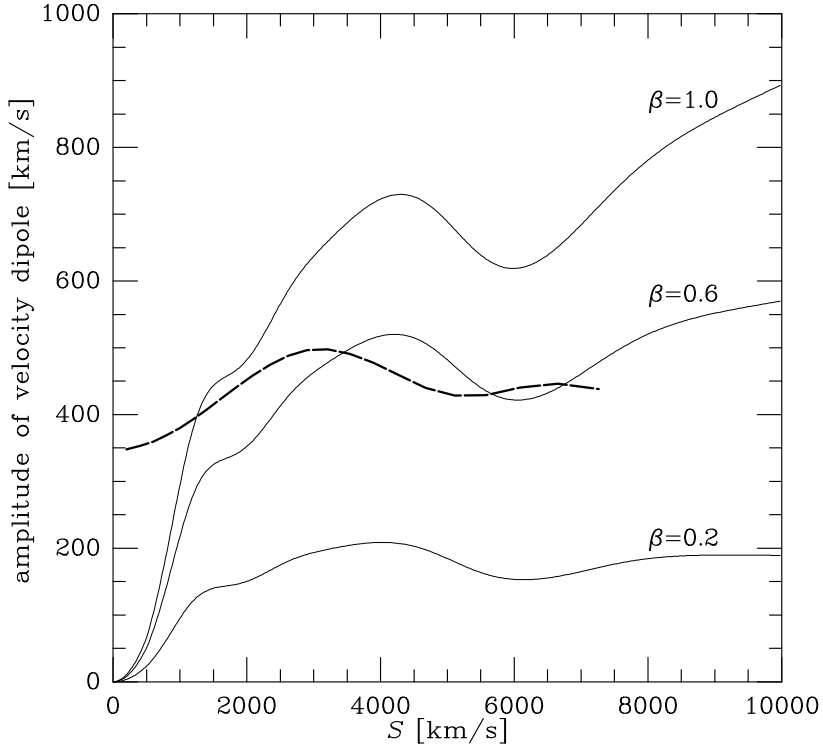

**Figure 1:** The light curves show the velocity dipole of the 1.2Jy IRAS catalog for $\beta = 0.2, 0.6$ and 1. The heavy curve is the dipole determined from peculiar velocity measurements as compiled by POTENT.

As discussed by Strauss *et al.* , high values of $\beta$ predict a large velocity dipole at large distances. Note that if the amplitude of the velocity is overestimated at some distant shell, the source term arising from the "rocket effect" correction will accordingly be too large, which will in turn cause the velocity to be even larger at more distant shell. In Figure 1 we see that $\beta \sim 0.7$ predicts a dipole of $\sim 600$ km s$^{-1}$ in accordance with the motion of the LG measured from the anisotropy of the CMB. If the motion of the LG relative to the CMB is generated by structure more distant than plotted in figure 1, the inferred $\beta$ would be lower still. This uncertainty in the length scale of structure responsible for our CMB motion is the traditional weakness of the dipole test.

A much more compelling use of the dipole is to compare the gravitational dipole of figure 1 with the observed peculiar velocity field. Plotted as the heavy curve in Figure 1 is the dipole in the LG frame inferred from the measured peculiar velocities of galaxies as processed in the most recent POTENT compilations (Dekel *et al.* 1993b). POTENT is a useful tool for this analysis, since it is capable of generating a full sky velocity map from an unevenly sampled sky distribution. The POTENT inferred dipole direction points only 11 degrees from the IRAS dipole, and the amplitude is consistent with the IRAS dipole for $\beta \sim 0.6$. The fact that the POTENT dipole does not approach 0 for small scale is an artifact of the 1200 km/s gaussian smoothing in the POTENT analysis; the average flow velocity of material within one smoothing scale of the origin does not exactly match the motion of the local group. Scales less than 2000 km/s should therefore be discounted for the intercomparison of POTENT and IRAS data in figure 1. The main source of error in this comparison arises from the rather large random errors associated with the measured velocities. Systematic errors such as inhomogeneous Malmquist bias are reasonably corrected for by the recent POTENT application and are not expected to have a major effect. Taking random errors in POTENT into account, we see from figure 1 that $\beta \sim 0.6 \pm 0.2$

## 4. DISCUSSION

Given the distribution of galaxies in redshift space, we have presented a self-contained method for recovering the associated velocity field. The method is based on a linear relation between the velocity and density and is fully expressed in redshift space, allowing us to estimate the velocity field without the necessity of iterating maps from redshift to real space. Clearly the method needs to be developed further, especially to include nonlinear effects which might be important for higher multipoles.

We have shown that the velocity dipole in the Local Group frame at a given shell is independent of external sources. This allows an unambiguous method of determination of $\beta = f(\Omega)/b$ by the comparison of the inferred velocity dipole from redshift surveys, *e.g.* IRAS, with the measured dipole of observed radial velocities. Our best estimate is $\beta = 0.6 \pm 0.2$, a factor of two smaller than reported for the POTENT-IRAS density field comparison (Dekel *et al.* 1993a). The reason for the discrepancy is not yet known, but the two analyses are quite independent.

Higher multipoles are of interest as well. Because the spherical harmonic multipoles of a Gaussian linear fluctuation field are independent variables, a detailed comparison between the observed velocity multipoles and all significant velocity multipoles determined from (12) should serve to further tighten the constraint on $\beta$. This comparison is currently under investigation and shall be reported in due course.

### Acknowledgments

We thank A. Dekel for stimulating discussions. We are especially thankful to the POTENT and IRAS teams for allowing us to use their data prior to publication. This work was supported in part by NSF grant AST-9221540 and NASA grant NAG-51360.